\documentclass[pra,twocolumn,showpacs,superscriptaddress,preprintnumbers,floatfix]{revtex4}
\usepackage{graphics}
\usepackage{amssymb}
\usepackage{color}
\usepackage[all]{xy}
\usepackage{graphicx}
\usepackage{caption}\CompileMatrices
\usepackage{dsfont}

\newtheorem{prop}{Proposition}\def\PRO{\begin{prop}}\def\ORP{\end{prop}}
\newtheorem{coro}{Corollary}\def\COR{\begin{coro}}\def\ROC{\end{coro}}
\newtheorem{theo}{Theorem}\def\TH{\begin{theo}}\def\HT{\end{theo}}
\def\TH{\begin{theo}}\def\HT{\end{theo}}
\newtheorem{defi}[prop]{Definition}\def\DE{\begin{defi}}\def\ED{\end{defi}}
\newtheorem{lemme}[prop]{Lemma}\def\LE{\begin{lemme}}\def\EL{\end{lemme}}

\def\EQ#1{\begin{eqnarray}#1\end{eqnarray}}

\def\bra#1{\langle#1{|}}

\def\ket#1{| #1 \rangle}

\def\bra#1{\langle #1 |}

\begin{document}

\title{Negative Conditional Entropy of Post-Selected States}

\author{Sina Salek}
\email[]{sina.salek@bristol.ac.uk}

\author{Roman Schubert}
\email[]{roman.schubert@bristol.ac.uk}

\author{Karoline Wiesner}
\email[]{k.wiesner@bristol.ac.uk}

\affiliation{School of Mathematics\\
The University of Bristol, Bristol BS8 1TW, United Kingdom}

\begin{abstract}

We define an quantum entropy conditioned on post-selection which has the von
Neumann entropy of pure states as a special case. This conditional entropy can
take negative values which is consistent with part of a quantum system 
containing less information than the whole which can be in a pure state. The definition is
based on generalised density operators for post-selected ensembles. 
The corresponding density operators are consistent with the quantum
generalisation of classical conditional probabilities following Dirac's
formalism of quasi-probability distributions. 

\end{abstract}

\pacs{03.65.Ta, 03.65.Wj, 03.65.Ca, 03.67.-a}
\maketitle

\section{Introduction \label{sec.prelim}}
Post-selection refers to keeping the record of the outcome of some ensemble quantum measurement 
only for those parts of the ensemble which at a later point in time 
are in a desired, so-called postselected state
and discarding the remaining results.
  Ensembles prepared in a state $\ket{\psi}$ and post-selected in a state
$\ket{\phi}$ are described by a generalised density operator as \cite{Reznik1995,Hosoya2010},
\EQ{
\rho_{\psi | \phi}:=\frac{\ket{\psi}\bra{\phi}}{\bra{\phi}\psi \rangle \label{W}}.
}
This generalisation is appropriate when a weak or no measurement has been performed between pre- and post- selection. For a proposal that incorporates strong measurements see \cite{Silva2014}. Nevertheless, we restrict our attention
to the case where no strong measurement has been performed. These generalised density operators for post-selected ensembles are 
used to obtain the so-called weak values $\Pi_w$
\cite{Aharonov1988} of an operator $\Pi$ as 
\EQ{
\Pi_w=Tr[\rho_{\psi | \phi} \Pi]. \label{weakval}
} 

Experimentally, a weak value is obtained by weakly coupling an ensemble of states
to a measuring apparatus, and post-selecting at a later time. 
To have an intuition about weak measurement on pre- and post- selected
ensembles, take the 3-box problem \cite{Aharonov1999}. Here, at time $t=0$ a
state $\ket{\psi}$ is
prepared in a superposition of states $\ket{A}$, $\ket{B}$ and $\ket{C}$ (the
three boxes),
\emph{e.g.} $\ket{\psi}=\frac{1}{\sqrt{3}}(\ket{A}+\ket{B}+\ket{C})$. At a
later time $t=1$ the system is weakly measured in the basis
$\{\ket{A},\ket{B},\ket{C}\}$ and then post-selected in some other state 
$\ket{\phi}$ which is not orthogonal to $\ket{\psi}$, \emph{e.g.}
$\ket{\phi}=\frac{1}{\sqrt{3}}(\ket{A}+\ket{B}-\ket{C})$. The resulting weak
values
of the operator projecting into the three boxes at time $t=1$ with
post-selection in state $\ket{\phi}$ at time $t=2$, calculated from Eq.
(\ref{weakval}),
are, respectively,
\EQ{\label{eq.ABC}
 A_w=1~,~~
 B_w=1~,~~
C_w=-1~.
}

These results can also be mathematically studied in the framework of Dirac quasi-probabilities. In 1945, Paul Dirac introduced a complex phase-space distribution to make an
``Analogy Between Classical and Quantum Mechanics"\cite{Dirac1945}, given by
\EQ{
\Pr(a_m,b_n)=\mbox{Tr}[\rho \Pi_A^m \Pi_B^n]~, \label{dirac}
}
where the $a_m$ and $b_n$ are the eigenvalues of the operators $\Pi_A$ and
$\Pi_B$, and $\Pi_A^m$ and $\Pi_B^n$ are the projectors onto the corresponding eigenstates. 

The Dirac distribution (\ref{dirac}) satisfies all the conditions of
classical Kolmogorov probabilities, except that it is not a positive real
function. It was shown that the negativity and complexness of this function is
due to the non-commutativity of the quantum mechanical observables
\cite{Johansen2007}. The Dirac distribution is normalised and gives correct marginals,
\EQ{
\sum_n \Pr(a_m,b_n)=\sum_n \mbox{Tr}[\rho A_m B_n]=\mbox{Tr}[\rho A_m]}
and
\EQ{
\sum_m \Pr(a_m,b_n)=\sum_m \mbox{Tr}[\rho A_m B_n]=\mbox{Tr}[\rho B_n],}
it obeys the sum rule and the product rule, and it is compatible with Bayes' law.  

Note, that 
Dirac distributions are not limited to phase space. In fact any two
operators with non-vanishing overlap between each of their eigenstates can be
used to construct a Dirac decomposition,

\EQ{\rho=\sum_{m,n}\Pr(a_m,b_n) \frac{\ket{a_m}\bra{b_n}}{\langle b_n\ket{a_m}}, \label{rhorec}} 

as long as the operators have the same Hilbert space dimension as the state
$\rho$, and their eigenvectors are mutually
non-orthogonal and none of those eigenvectors are orthogonal to the 
state $\rho$. 
Hence a space spanned by any two such observables would be
sufficient to describe all the information available from the state $\rho$.
This is due to the fact that one can
describe any quantum state of $d$-dimensional Hilbert Space with $d^2-1$
elements. 

 For theoretical
considerations on the Dirac distribution see the work by
Johansen~\cite{Johansen2007} and Hofmann \cite{Hofmann2012}. An experimental
procedure for measuring the Dirac distribution of a general quantum state has
been given by Lundeen and Bamber~\cite{Lundeen2012}. 

We will now describe how weak values  can be understood in the framework
of Dirac distributions.

Note that we can rewrite the weak value, Eq.~(\ref{weakval}), as 
\EQ{
\Pi_w=\frac{\bra{\phi}\Pi \ket{\psi}}{\langle\phi\mid\psi\rangle} \label{weak}.
}
Now we can interpret $\Pi_w$ as a conditional Dirac distribution of an ensemble which is pre- and
post-selected in states 
$\ket{\psi}$ and $\ket{\phi}$, respectively.
Using Bayes' law and defining $\rho_\psi:=\ket{\Psi}\bra{\Psi}$ and $
\rho_\phi:=\ket{\Phi}\bra{\Phi}$, the weak value 
can be written as 
\EQ{
\Pi_w=\Pr(\Pi|\phi) = \frac{\mbox{Tr}[\rho_\psi \rho_\phi \Pi]}{\mbox{Tr}[\rho_{\psi} \rho_{\phi}] \label{conditional}}.
}
In this last equation the interpretation as a conditional
quasi-probability follows from the use of Bayes' law.

This close connection between weak values and Dirac quasi-probabilities gives
an operational meaning to the complex values of the Dirac distribution as a
result of  weak measurements. In the case where the measurement was performed by coupling the momentum of the measurement pointer to the quantum system, the real part of the weak
value refers to the shift in the position of the measurement pointer, while
the imaginary part refers to the shift in the momentum of the measurement
pointer \cite{Jozsa2007}. The same interpretation can be given to the
real and complex part of the Dirac distribution. Given this operational
meaning of the Dirac distribution, we proceed to use this formalism to make an
analogy between classical conditional probabilities and quantum conditional
states.

In the following, we establish first that Eq.~(1) is actually a form of a conditional state
by using the framework of Dirac distributions. 
This allows us then to define a corresponding conditional entropy of
post-selected quantum states.

\section{Quantum \emph{conditional} states and \emph{conditional} Entropy}

\subsection{Quantum states, conditioned on post-selection}

Using the conditional Dirac distribution obtained in Eq. (\ref{conditional}), we 
now construct the corresponding conditional Dirac decomposition $\rho_{\psi|\phi}$
in analogy to Eq.~(\ref{rhorec}), given post-selection in some state $\ket{\phi}$. 

The summation here runs over the
eigenstates of a projection operator $\Pi$ measured weakly in between the times of
pre-selection and post-selection.
We define $\Pi:=\ket{h}\bra{h}$, where $\ket{h}$ is one out of a complete set
of basis states $\{\ket{h}\}$ which are mutually non-orthogonal to the
state $\ket{\phi}$.\footnote{We use the
letter $h$ for the basis states of the projection operator to honour the
closeness of these concepts to the consistent histories approach of quantum
measurement, see \cite{Hartle2008}.}

The conditional Dirac decomposition $\rho_{\psi|\phi}$ we obtain is 
\EQ{
\rho_{\psi|\phi} :=\sum_{\rm h} \Pr({\rm h}|\phi) \frac{\ket{\rm h}\bra{\phi}}{\langle \phi\ket{\rm h}}. \label{condi}
}
$\rho_{\psi|\phi}$ is the density operator of 
all paths leading from state $\ket{\psi}$ to  post-selected state $\ket{\phi}$. 
One may note that the operator $\rho_{\psi|\phi}$ is not Hermitian. It has been argued in the weak measurement literature why this should not be a cause of concern \cite{Lundeen2012}.
$\rho_{\psi|\phi}$ thus defined is a trace-one operator and can indeed, by
construction, be determined by weak measurements. 
In the case of the three-box problem, the eigenstates
$\ket{h}$ of the projector $\Pi$ would be $\ket{A}$, $\ket{B}$, and $\ket{C}$,
representing the system in being in one of the three boxes $A$, $B$, and $C$
at time $t=1$. 
 
By writing  
\EQ{ \Pr({\rm h}|\phi)= \frac{ \bra{\phi}{\rm h}\rangle\bra{\rm
h}\psi\rangle}{ \bra{\phi}\psi\rangle}
}
and inserting it into  Eq.~(\ref{condi}) we obtain
\EQ{
\nonumber \rho_{\psi|\phi}
\nonumber &=&\sum_{\rm h} \frac{\ket{\rm h}\bra{\rm h}\psi\rangle\bra{\phi}}{\bra{\phi}\psi\rangle}\\
&=&\frac{\ket{\psi}\bra{\phi}}{\bra{\phi}\psi\rangle} \label{condi2},
}
which is indeed the generalised density operator for post-selected ensembles, defined in Eq.~(\ref{W}).
This links the interpretation 
Eq.~(\ref{condi}) as conditional quantum states to
the generalised density operator for post-selected ensembles as in Eq.~(\ref{W}).

We observe that the application of Bayes' law in the definition
(\ref{condi}) results in a density operator which is the extension of
classical conditional probabilities. 
This analogy can be further clarified by multiplying $\rho_{\psi|\phi}$ by
probability $\Pr(\phi)$ of the system ending in state $\ket{\phi}$ and sum
Eq.~(\ref{condi}) over a complete basis $\{\ket{\phi}\}$. Thus, we retrieve the density operator of Eq.~(\ref{rhorec}) which contains the full information about the system,
\EQ{\rho=\sum_\phi \Pr(\phi) \rho_{\psi|\phi} , \label{sum}}
with $\Pr(\phi)=Tr[\rho \ket{\phi} \bra{\phi}]$ being the probability of the state being in the state $\ket{\phi}$ at the time of post-selection. This is what is expected from the analogy with classical probabilities.

\subsection{Entropies, conditioned on post-selection }

In classical information
theory, the entropy of random variable $X$ conditioned on selection of a
particular instance $y$ of random variable $Y$ is $H(X|Y=y)=-\sum_x \Pr(x|y)
\log \Pr(x|y)$, where $\Pr(x|y)$ is the conditional probability of $x$
given the occurrence of $y$. Note that this function is the average 
$\Big\langle - \log \Pr(x|y)\Big\rangle$.  The classical  conditional entropy
is, again, an average, 
i.e. $H(X|Y)=\sum_y \Pr(y)H(X|Y=y)$.

In analogy to the classical conditional entropy, we define the conditional
entropy of a state $\ket{\psi}$, given 
post-selection in one out of a possible set of states $\ket{\phi}$, to be
\EQ{
S_c(\psi|\Phi=\phi)=-\frac{1}{2}\mbox{Tr}[\rho_{\psi|\phi}\log(\rho^{ }_{\psi|\phi}\rho^\dagger_{\psi|\phi})], \label{CS}
}
where $\rho^\dagger_{\psi|\phi}$ is the conjugate transpose of
$\rho_{\psi|\phi}$. We note that this particular form of the entropy is an
average of the logarithm of the density operator of the system: \begin{equation}S_c(\psi|\Phi=\phi) = \Big\langle - \log
\Big((\rho^{
}_{\psi|\phi}\rho^\dagger_{\psi|\phi})^{1/2}\Big)\Big\rangle~.\end{equation}

One could choose other functional forms for the conditional entropy here. For
instance, one could choose the function $-\frac{1}{2}\mbox{Tr}[\rho^{
}_{\psi|\phi}\rho^\dagger_{\psi|\phi}{\log}(\rho^{
}_{\psi|\phi}\rho^\dagger_{\psi|\phi})]$  instead. However, this would not
correspond to an average of the logarithm.  
This makes our  choice for the conditional entropy function a natural one. 
Note that the argument of the logarithm gives the Singular values of the operator $\rho_{\psi|\phi}$.
The Singular Value Decomposition is a generalisation of diagonalisation needed
here 
due to the particular choice of basis in the definition of the two-state density
operator (Eq.~\ref{condi}). For Hermitian
operators, such as ordinary density operators, the singular values and the eigenvalues are the same.

Finally, we define the general quantum conditional entropy as an average of
the entropy conditioned on a particular choice of post-selection as
\EQ{S_C(\psi|\Phi)=\sum_\phi \Pr(\phi)S_c(\psi|\Phi=\phi) \label{CS2}.}

Eqs.~(\ref{CS} -- \ref{CS2}) can be simplified. By inserting Eq.~(\ref{W}) into Eq.~(\ref{CS}) we obtain 
\EQ{ 
\nonumber S_c(\psi|\Phi=\phi)&=&-\frac{1}{2}\frac{\bra{\phi}{\log}(\rho^{ }_{\psi|\phi}\rho^\dagger_{\psi|\phi})\ket{\psi}}{\bra{\phi}\psi\rangle}\\
  &=&{\log}|\bra{\phi}\psi\rangle|~.
}

Inserting this into Eq.~(\ref{CS2}) we obtain the simplified expression 
\EQ{
S_C(\psi|\Phi)= \sum_\phi |\bra{\phi}\psi\rangle|^2
{\log}|\bra{\phi}\psi\rangle|~.
}
This new expression is very useful for finding upper and lower bounds of the conditional
entropy.  
We can see that $S_C$ has a negative lower bound of
$\frac{1}{d}\log\frac{1}{\sqrt{d}}$, where $d$ is the dimension of the  Hilbert
Space. This lower bound is reached when all states $\ket{\phi}$ have the same
magnitude in overlap with the state $\ket{\psi}$. 
 $S_C$ is bounded from above by the von Neumann entropy of the system
without post-selection. 

The conditional entropy $S_C$ can be understood as the amount of information
contained in a system 
conditioned on a particular post-selection. The case discussed here is
the  special case of pre-selection in a pure state which, by definition, has
zero von Neumann entropy. The upper bound of the conditional entropy is
zero in this case and thus the 
conditional entropy of a pure pre-selected state $\ket{\psi}$ can be
negative. This is a necessary consequence from  choosing a subset of the entire
(pure) system for post-selection which necessarily
decreases the entropy below zero. 
Thus, a negative conditional entropy of post-selected ensembles becomes 
intuitive as a subset contains less information than the whole. The concept of
negative conditional entropy appears in other contexts in quantum information.
For instance, for a bipartite system $\rho_{AB}$ in a pure, i.e. maximally
entangled state, the joint entropy is zero. On the other hand, the locally
the subsystems $\rho_A$ and $\rho_B$ are in maximally mixed states.
Hence, the entropy of one subsystem, conditioned on the measurement of the
other, $S(\rho_A|\rho_B)=S(\rho_{AB})-S(\rho_{B})<0$, takes on a negative
value. This conditional entropy was also calculated in the form of
$S(\rho_A|\rho_B)=-\mbox{Tr}[\rho_{AB} {\log}\rho_{A|B}]$, where
$\rho_{A|B}=\lim_{n \to +\infty}[\rho_{AB}^{1/n}(\mathds{1}_A \otimes
\rho_B)^{-1/n}]^n$ \cite{Cerf1996} which is closer in outlook to the 
construction introduced in Eq.~(\ref{CS2}).


\paragraph{Example: 3-box post-selection}

We will now revisit the 3-box problem \cite{Aharonov1999} from the beginning and calculate the
conditional entropies just defined. 

Choosing pre- and post-selection as before, i.e. 
$\ket{\psi}=1/\sqrt{3}(\ket{A}+\ket{B}+\ket{C})$ and 
$\ket{\phi}=1/\sqrt{3}(\ket{A}+\ket{B}-\ket{C})$, one obtains for 
the conditional entropy $S_c$ of the system, Eq.~(\ref{CS}),
$S_c(\psi|\Phi=\phi)=-\ln 3$. In order  to calculate the
conditional entropy $S_C$, Eq.~(\ref{CS2}), given that the Hilbert space
dimension of the system is three, one needs to perform the same calculation as
above for two other post-selected states, $\ket{\phi'}$ and $\ket{\phi''}$.
The only restrictions on the choice of these two post-selected states are that
they need to be mutually non-orthogonal to the state of each box, i.e. to
$\ket{A}$, $\ket{B}$ and $\ket{C}$. And in addition they need, together with our original state $\ket{\phi}$, to span the Hilbert space of the state $\ket{\psi}$. Take these states to be 
\EQ{
\ket{\phi'}=\frac{1}{\sqrt{3}} \ket{A}+\frac{-3-\sqrt{3}}{6}\ket{B}+\frac{-3+\sqrt{3}}{6}\ket{C}
}
and
\EQ{
\ket{\phi''}=\frac{1}{\sqrt{3}} \ket{A}+\frac{3-\sqrt{3}}{6}\ket{B}+\frac{3+\sqrt{3}}{6}\ket{C}.
}
Calculating $S_c$ of Eq.~(\ref{CS}) for
for $\ket{\phi'}$ and $\ket{\phi''}$
gives $S_c(\psi|\Phi=\phi')=-\log_3 4.10$ and $S_c(\psi|\Phi=\phi)=-\log_3
1.10$. 
With probabilities $\Pr(\phi)=0.11$, $\Pr(\phi')=0.06$ and $\Pr(\phi'')=0.83$
(given by $|\bra{\phi}\psi\rangle|^2$ and correspondingly for $\phi^\prime$
and $\phi^{\prime\prime}$), 
the conditional entropy Eq.~(\ref{CS2}) becomes $S_C(\psi|\Phi)=-0.26$, with
the logarithm being calculated in base 3 for convenience.

The 3-box problem illustrates the negativity of the conditional entropy where
each path through a box contains partial information of the system as a whole.
 

\paragraph{Conclusion}

In this work we have defined an entropy for post-selected ensembles. To this
end, we have adapted the formalism of Dirac quasi-probabilities, and showed
that the two-state density operator of post-selected ensembles is a
generalisation of the classical conditional probability, given that a
particular final state is selected. We have found upper and lower bounds on
such entropies and interpreted the values as the amount of information
contained in a system conditioned on a particular post-selection. Furthermore, we
showed that these states have properties beyond their classical
counterparts. Most notably, conditional quantum entropies as defined in this work can have negative values.
This is consistent with our interpretation. This new quantum entropy should
open up avenues for studying the properties of weak measurement of more
general states than discussed in this paper, such as mixed post-selected
states.

{\bf Acknowledgements} We thank Jeff Lundeen, Holger Hofmann and Sandu
Popescu for their helpful comments and stimulating discussions. K.W. thanks EPSRC for financial support.

{\footnotesize

\bibliography{e012119}
\bibliographystyle{plain}

}

\end{document}